\documentclass[11pt]{article}
\usepackage{longtable}
\usepackage{ae,aecompl,pslatex}
\usepackage[dvips]{graphicx}
\usepackage{enumerate,color}
\usepackage{fullpage}
\usepackage{lineno}

\usepackage{placeins}
\usepackage{psfrag,epsfig,graphicx,color,array,dcolumn,tabularx,delarray,longtable,indentfirst,amsmath,amsfonts,amssymb,amsthm,eucal,bbm}
\theoremstyle{bkaexa} 
\usepackage{caption}
\usepackage{setspace}
\newtheorem{Exa}{Example}
\theoremstyle{bkaexa} 
\newtheorem{Rem}{Remark}
\theoremstyle{bkathm} 

\theoremstyle{bkathm} 

\theoremstyle{bkathm} 

\theoremstyle{bkathm} 

\theoremstyle{definition}

\begin{document}
\setstretch{1.5}
\title{A note on distance variance for categorical variables}
\author{\normalsize Qingyang Zhang\\
\normalsize Department of Mathematical Sciences, University of Arkansas, AR 72701\\
\normalsize Email: qz008@uark.edu
}
\date{}
\maketitle

\begin{abstract}
This study investigates the extension of distance variance, a validated spread metric for continuous and binary variables [Edelmann et al., 2020, \textit{Ann. Stat.}, 48(6)], to quantify the spread of general categorical variables. We provide both geometric and algebraic characterizations of distance variance, revealing its connections to some commonly used entropy measures, and the variance-covariance matrix of the one-hot encoded representation. However, we demonstrate that distance variance fails to satisfy the Schur-concavity axiom for categorical variables with more than two categories, leading to counterintuitive results. This limitation hinders its applicability as a universal measure of spread.
\end{abstract}


\section{Introduction}
Quantifying the spread of data is a fundamental task in statistics and various scientific disciplines. For continuous variables, well-established measures like variance, trimmed variance, and interquartile range (IQR) are readily available. However, measuring spread for categorical (nominal) variables poses a unique challenge, because these variables lack inherent order or a direct numerical representation. Traditional approaches for categorical variables, such as the Gini index, focus on the level of uncertainty associated with random sampling. Newer measures based on penalty functions are more attractive, as they can be interpreted in terms of prediction \cite{Gilula, Haberman}. Despite these advancements, a universal measure of spread that seamlessly applies to both continuous and categorical variables remains elusive.

Since the seminal work of Sz\'{e}kely et al. (2007), distance correlation analysis has gained popularity among statisticians and practitioners. This correlation metric is distribution-free and can handle diverse variable types, including categorical variables and random vectors \cite{dcor, Shenetal, Zhang2019}. Notably, distance correlation shares a conceptual similarity with Pearson's correlation. It can be expressed as distance covariance normalized by distance variance. Edelmann et al. (2020) explored the distance variance as a potential spread measure for continuous and binary data \cite{DStd}. Importantly, they demonstrated that distance variance satisfies Bickel \& Lehmann's three axioms \cite{Bickel}, solidifying its potential as a valid measure of spread. Motivated by the pursuit of a universal spread metric, this paper investigates the applicability of distance variance to quantify the spread of general categorical variables. 

We begin with the mathematical expressions of this metric. Let $X$ be a categorical variable taking values from $\{1,~...,~K\}$ with probability mass function $P(X = k) = \pi_{k}$, where $\pi_{k}>0$ and $\sum_{k=1}^{K}\pi_{k} = 1$. A one-hot encoded representation of $X$ is $\mathbf{Z} = (I\{X = 1\}, ..., I\{X = K\})^{T}$. The distance variance of $X$ based on Euclidean distance, i.e., $d(x_{1},~x_{2}) := |\mathbf{z}_{1} - \mathbf{z}_{2}|_{2} = 0$ if $x_{1} = x_{2}$ and $\sqrt{2}$ otherwise, can be expressed as follows (\cite{Zhang2019}, eq. 3)
\begin{align}
\mathcal{V}(X) & =  \left\{E[d^{2}(X_{1},~ X_{2})] + E^{2}[d(X_{1},~ X_{2})] - 2E[d(X_{1},~ X_{2})d(X_{1},~X_{3})] \right\}^{\frac{1}{2}} \nonumber \\
& = \sqrt{2} \left[ \left( \sum_{k=1}^{K}\pi_{k}^2 \right)^2 + \sum_{k=1}^{K}\pi_{k}^2 - 2\sum_{k=1}^{K}\pi_{k}^3 \right]^{\frac{1}{2}},
\end{align}
where $(X_{1},~X_{2},~X_{3})$ are three independent copies of $X$. Interestingly, for categorical variables, distance variance is invariant (up to a constant) of the choice of distance metrics. For instance, the $\alpha$-distance variance \cite{dcor} can be expressed as
\begin{align*}
\mathcal{V}_{\alpha}(X) & =  \left\{E[d^{2\alpha}(X_{1},~ X_{2})] + E^{2}[d^{\alpha}(X_{1},~ X_{2})] - 2E[d^{\alpha}(X_{1},~ X_{2})d^{\alpha}(X_{1},~X_{3})] \right\}^{\frac{1}{2}}\\
& = 2^{\frac{\alpha}{2}} \left[ \left( \sum_{k=1}^{K}\pi_{k}^2 \right)^2 + \sum_{k=1}^{K}\pi_{k}^2 - 2\sum_{k=1}^{K}\pi_{k}^3 \right]^{\frac{1}{2}}.
\end{align*}

The distance variance based on the Gaussian kernel with bandwidth $\sigma^2$, i.e., $d(x_{1},~x_{2}) := \kappa(\mathbf{z}_{1}, \mathbf{z}_{2}) = \exp(-|\mathbf{z}_{1} - \mathbf{z}_{2} |_{2}/\sigma^2)$, is also equivalent to $\mathcal{V}(X)$ 
\begin{align*}
\mathcal{V}_{\kappa}(X, \sigma^{2}) & =  \left\{E[\kappa^{2}(\mathbf{Z}_{1},~ \mathbf{Z}_{2})] + E^{2}[\kappa(\mathbf{Z}_{1},~ \mathbf{Z}_{2})] - 2E[\kappa(\mathbf{Z}_{1},~ \mathbf{Z}_{2})\kappa(\mathbf{Z}_{1},~\mathbf{Z}_{3})] \right\}^{\frac{1}{2}}\\
 & = \left( 1- e^{-\frac{\sqrt{2}}{\sigma^2}} \right) \left[ \left( \sum_{k=1}^{K}\pi_{k}^2 \right)^2 + \sum_{k=1}^{K}\pi_{k}^2 - 2\sum_{k=1}^{K}\pi_{k}^3 \right]^{\frac{1}{2}}.
\end{align*}

More generally, it can be shown that for any two constants $c_{1}\neq c_{2}$, such that $|X_{1}-X_{2}| = c_{1}$ if $X_{1} = X_{2}$ and $c_{2}$ otherwise, the distance variance is equivalent to $\mathcal{V}(X)$ up to a constant (proof is provided in the Appendix). Throughout this paper, we refer to $\mathcal{V}(X)$ based on Euclidean distance, unless otherwise stated. Given a sequence of independent and identically distributed (i.i.d.) samples of $X$, an unbiased estimate of $\mathcal{V}^{2}(X)$ can be computed as follows (\cite{Zhang2024}, page 14)
\begin{align*}
\widehat{\mathcal{V}^{2}(X)} = & \frac{n^3}{(n-1)(n-2)(n-3)}\left( 1-\sum_{k = 1}^{K}\widehat{\pi}_{k}^{2} \right)^2 -\frac{2n^2}{(n-2)(n-3)} \sum_{k = 1}^{K}\widehat{\pi}_{k}^{3}  \\
& - \frac{n(n-6)}{(n-2)(n-3)}\sum_{k = 1}^{K}\widehat{\pi}_{k}^{2} - \frac{n(n+2)}{(n-2)(n-3)},
\end{align*}
where $n$ is the sample size and $\widehat{\pi}_{k}$ is the maximum likelihood estimate of $\pi_{k}$. The estimate $\widehat{\mathcal{V}^{2}(X)}$ is essentially a fourth-order U-statistics, which is asymptotical normal. Standard methods for U-statistics such as Jackknifing can be directly applied for hypothesis testing purpose \cite{arvesen}. 

While the validity and usefulness of distance variance as a spread metric is well-established for continuous and binary variables, its suitability for multi-category variables (i.e., with $K\geq 3$ categories) remains unexplored. One challenge lies in interpreting $\mathcal{V}(X)$, which involves three independent samples $(X_{1},~X_{2},~X_{3})$. This complexity makes a straightforward interpretation difficult. This paper addresses two key questions:
\begin{itemize}
\item[\textbf{Q1}. ] How can we interpret distance variance for general categorical variables in intuitive ways?
\item[\textbf{Q2}. ] Is distance variance a valid measure of spread for general categorical variables?
\end{itemize} 
For Q1, we propose alternative interpretations of distance variance using matrix norms and inter-point distances, offering a more intuitive understanding. Unfortunately, for Q2, we show that distance variance is not a valid measure of spread for variables with more than two categories, and its misuse can lead to counterintuitive results.

The remainder of this paper is structured as follows: Section 2 characterizes distance variance and proposes a broad class of spread measures based on these characterizations. Section 3 first introduces a set of axioms, then demonstrates that distance variance violates the condition of strict Schur-concavity. Section 4 discusses the findings and concludes the paper.

\section{Geometric and algebraic characterizations for distance variance}
In this section, we give two intuitive interpretations of distance variance. These interpretations reveal important connections between the distance variance $\mathcal{V}(X)$, the variance-covariance matrix of $\mathbf{Z}$ and some popular entropy measures. We further propose a class of similar measures with potential applications as spread metrics. The validity of these proposed measures is discussed in Section 3. For clarity, we denote the $L_{p}$-norm for vectors by $|\cdot|_{p}$ and for matrices by $\|\cdot\|_{p}$ throughout the paper.

\subsection{Geometric characterization}
For notational convenience, we define $\mathbf{Z}(k)$ as the one-hot encoded representation of $X=k$, i.e., $\mathbf{Z}(k)$ is a vector with 1 in its $k$th position and 0 elsewhere. For example, when $K=3$, $\mathbf{Z}(1) = (1,~0,~0)$, $\mathbf{Z}(2) = (0,~1,~0)$, $\mathbf{Z}(3) = (0,~0,~1)$. The mean vector, $E(\mathbf{Z}) = (\pi_{1},~...,~\pi_{K})$, can be considered as the geometric center of $[\mathbf{Z}(1),~...,~\mathbf{Z}(K)]$. Interestingly, we find that the squared distance variance, $\mathcal{V}^{2}(X)$, is closely related to the distances between each ${Z}(k)$ and the geometric center $E(\mathbf{Z})$. Specifically, it can be expressed as 
\begin{equation}
\mathcal{V}^{2}(X) = 2\sum_{k=1}^{K}\pi^{2}_{k}|\mathbf{Z}(k)-E(\mathbf{Z})|_{2}^{2}.
\end{equation}
\begin{proof}
\begin{align*}
\sum_{k=1}^{K}\pi^{2}_{k}|\mathbf{Z}(k)-E(\mathbf{Z})|_{2}^{2} & = \sum_{k=1}^{K}\pi^{2}_{k}\left[ (1-\pi_{k})^{2} + \sum_{k'=1}^{K}\pi^{2}_{k'} -  \pi^{2}_{k} \right] \\
& = \sum_{k=1}^{K}\pi^{2}_{k}(1-\pi_{k})^{2} + \left(\sum_{k=1}^{K}\pi^{2}_{k}\right)^{2} - \sum_{k=1}^{K}\pi^{4}_{k} \\
& = \left(\sum_{k=1}^{K}\pi^{2}_{k}\right)^{2}  + \sum_{k=1}^{K}\pi^{2}_{k} - 2\sum_{k=1}^{K}\pi^{3}_{k} \\
& = \frac{1}{2}\mathcal{V}^{2}(X).
\end{align*}
\end{proof}

Equation 2 reveals that the squared distance variance can be interpreted as a weighted average of the squared distances between each category and the geometric center. Figure 1 illustrates this for two specific cases, $K=2$ and $K=3$.

\begin{figure}[!htbp]
\begin{center}
\includegraphics[scale=0.3]{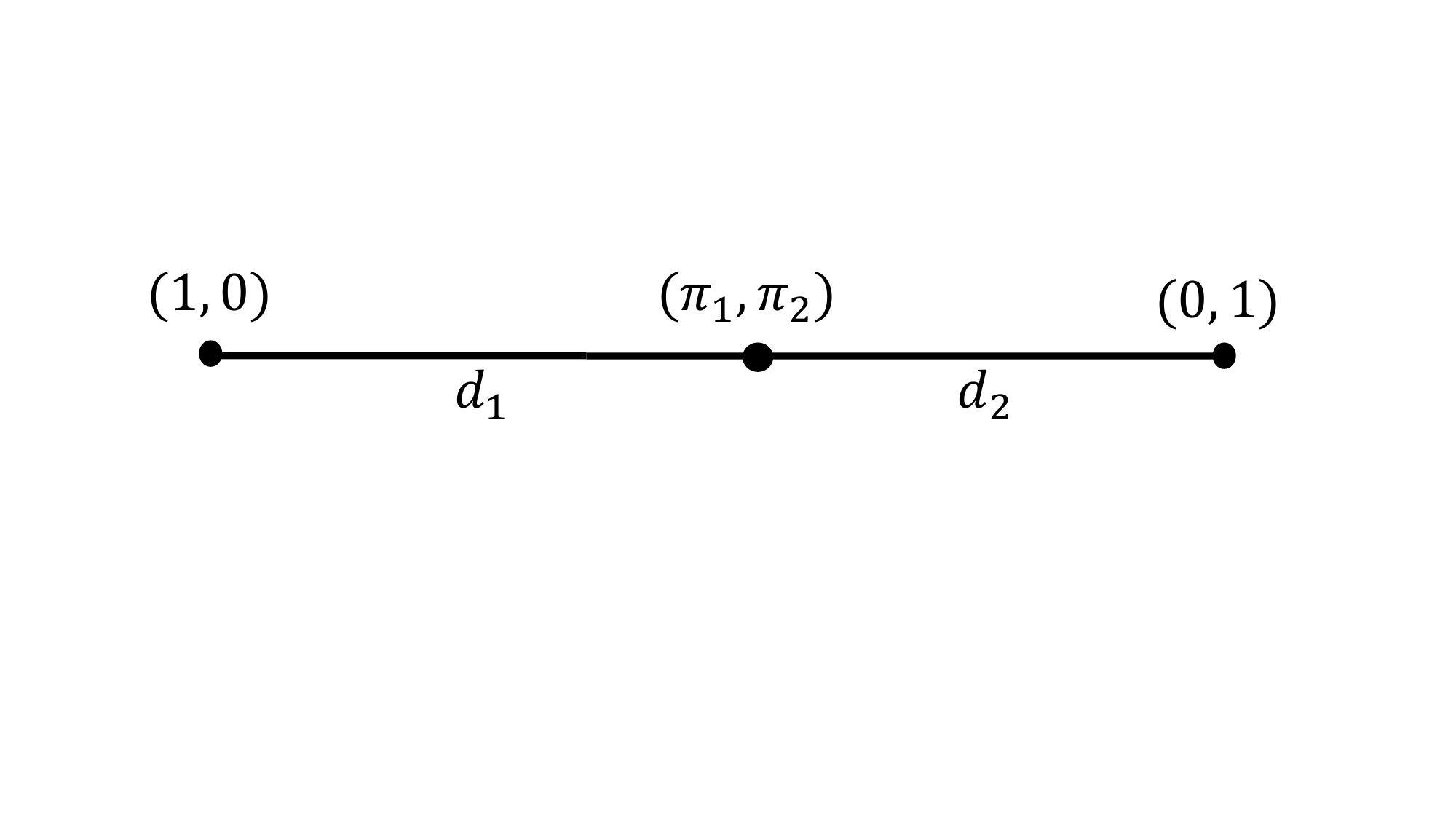}
\includegraphics[scale=0.3]{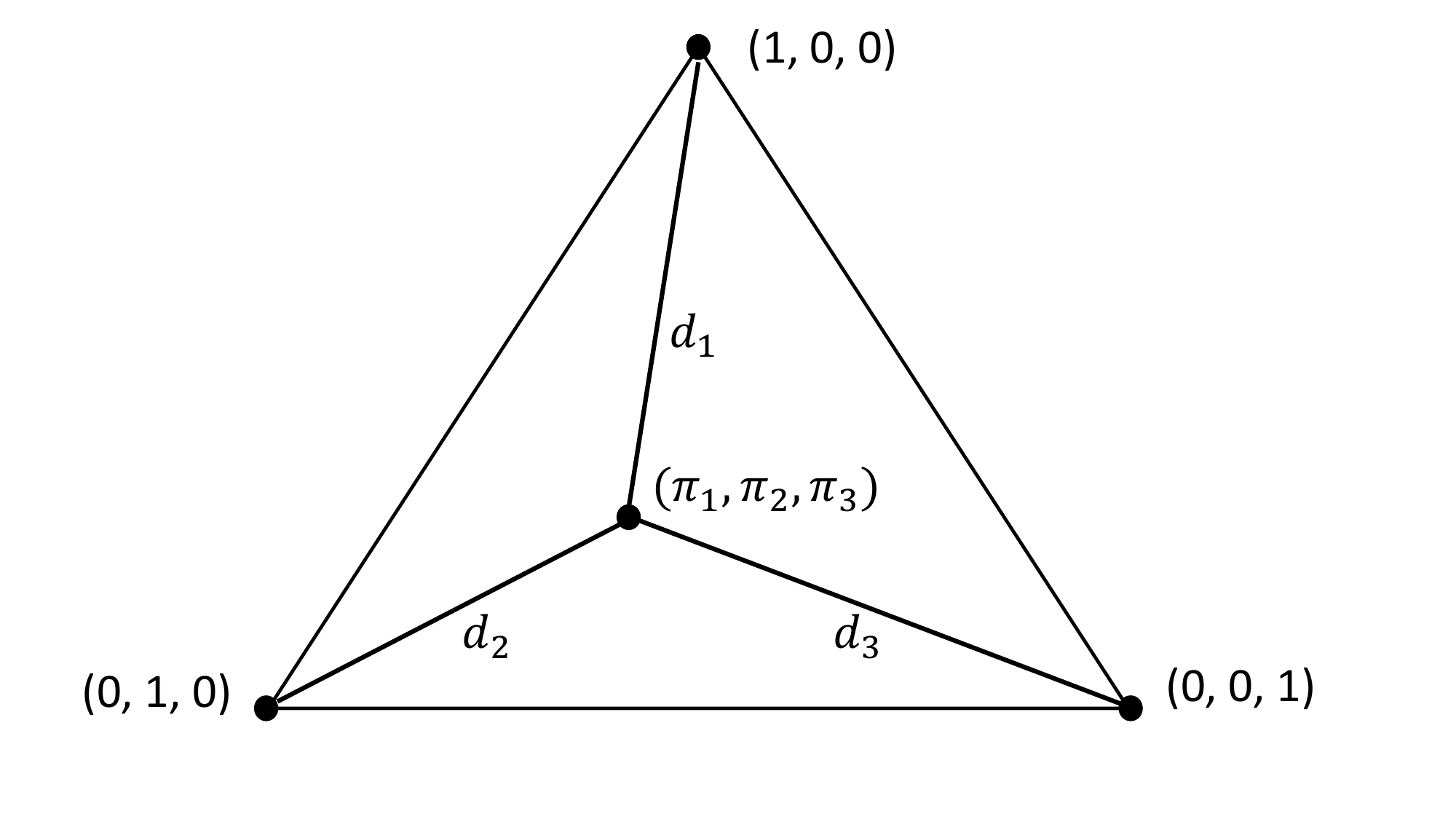}
\end{center}
\caption{Geometric interpretations of distance variance for $K=2$ and $K=3$, where $d_{k}$ represents the Euclidean distance between $\mathbf{Z}(k)$ and $E(\mathbf{Z})$.}
\end{figure}

Motivated by the geometric interpretation of $\mathcal{V}^{2}(X)$, we propose a generalized version 
\begin{equation*}
\Delta_{g}(X,~w,~l,~p) = \sum_{k=1}^{K}w(\pi_{k})|\mathbf{Z}(k)-E(\mathbf{Z})|_{p}^{l},
\end{equation*}
where the subscript $g$ denotes the geometric extension. The squared distance variance is a special case with $p=2$, $l=2$, $w(\pi_{k}) = \pi^{2}_{k}$. Another special case is the Gini index (also known as R\'{e}nyi entropy), $G(X) = 1-\sum_{k=1}^{K}\pi_{k}^{2}$, when $p=1$, $l=1$ and $w(\pi_{k}) = \pi_{k}$ or $w(\pi_{k}) = 1 + \pi_{k}$. 

Furthermore, for $p=1$, $l=1$, and a specific weighting function $w(\pi_{k}) = 1/(m-1)\cdot\sum_{q=0}^{m-1}\pi^{q}_{k}$, where $m>1$, $\Delta_{g}(X,~w,~l,~p)$ coincides with Tsallis entropy \cite{Tsallis}, defined as 
\begin{equation*}
H_{m}(X) = \frac{1}{m-1}\left( 1-\sum_{k=1}^{K}\pi_{k}^{m} \right).
\end{equation*}

Notably, Tsallis entropy encompasses several well-known entropy measures. This includes the Gini index $H_{2}(X)$ and Shannon entropy, obtained as the limit of Tsallis entropy when $m$ approaches 1, $H_{1}(X) := \lim_{m\rightarrow 1}H_{m}(X)= -\sum_{k=1}^{K}\pi_{k}\log \pi_{k}$. Another intriguing special case of $\Delta_{g}(X,~w,~l,~p)$ is the extropy, defined as $J(X) -\sum_{k=1}^{K}(1-\pi_{k})\log (1-\pi_{k})$. Extropy acts as the complementary dual of Shannon entropy (see Lad et al., 2015 for a detailed discussion on entropy and extropy). Within our framework, it's straightforward to verify that $J(X)$ aligns with $\Delta_{g}(X,~w,~l,~p)$ when $p=1$, $l=1$ and $w(\pi_{k}) = -\log(1-\pi_{k})$. 

In addition to these established entropy measures, our framework allows for the construction of many new metrics through various combinations of $l$, $p$ and $w$. The general validity of $\Delta_{g}(X,~w,~l,~p)$ will be explored in Section 3.

\subsection{Algebraic characterization}
A key distinction between distance variance and existing measures such as Tsallis entropy and extropy lies in the functional expression. Distance variance incorporates terms like $\pi_{k}\pi_{k'}$ (where $k\neq k'$) that arise in the first term of Equation 1, while other measures rely solely on marginal probabilities $\pi_{k}$. Note that $\mbox{Cov}(I\{X = k\},~I\{X = k'\} ) = -\pi_{k}\pi_{k'}$ for any $k\neq k'$. This suggests a strong connection between distance variance and the variance-covariance matrix of the one-hot encoded representation, given by
\begin{equation*}
V(\mathbf{Z}) = \begin{bmatrix} 
    \pi_{1}(1-\pi_{1}) & -\pi_{1}\pi_{2} & \dots & -\pi_{1}\pi_{K}\\
     -\pi_{2}\pi_{1} & \pi_{2}(1-\pi_{2}) & \dots & -\pi_{2}\pi_{K} \\
    \vdots & \vdots  & \ddots & \vdots \\
    -\pi_{K}\pi_{1} &  -\pi_{K}\pi_{2} & \dots & \pi_{K}(1-\pi_{K}) 
    \end{bmatrix}.
 \end{equation*}   
 
 Furthermore, the squared covariance terms in $\mathcal{V}^{2}(X)$ suggest a close relationship between distance variance and the $L_2$-norm (Frobenius norm) of $V(\mathbf{Z})$. In fact, a direct relationship can be established 
 \begin{equation}
 \mathcal{V}(X) = \sqrt{2}\|V(\mathbf{Z})\|_{2}.
 \end{equation}
\begin{proof}
\begin{align*}
\|V(\mathbf{Z})\|^{2}_{2} & = \sum_{k=1}^{K}\pi_{k}^{2}(1-\pi_{k})^{2} + \sum_{k=1}^{K}\sum_{k'\neq k}\pi_{k}^{2}\pi_{k'}^{2} \\
& = \sum_{k=1}^{K}\pi_{k}^{2}(1-\pi_{k})^{2} + \left(\sum_{k=1}^{K}\pi_{k}^{2}\right)^{2} - \sum_{k=1}^{K}\pi_{k}^{4} \\
& = \left( \sum_{k=1}^{K}\pi_{k}^2 \right)^2 + \sum_{k=1}^{K}\pi_{k}^2 - 2\sum_{k=1}^{K}\pi_{k}^3 \\
& =  \frac{1}{2}\mathcal{V}^{2}(X).
\end{align*}
\end{proof}

Equation 3 reveals that distance variance is equivalent to the $L_2$-norm of the variance-covariance matrix. Naturally, one may consider a possible extension using $L_p$ norms of $V(\mathbf{Z})$
\begin{equation*}
\Delta_{a}(X, p) = \|V(\mathbf{Z})\|_{p} = \left\{ \sum_{k=1}^{K}\pi_{k}^{p}(1-\pi_{k})^{p} + \left[ \sum_{k=1}^{K}\pi_{k}^{p} \right]^{2} - \sum_{k=1}^{K}\pi_{k}^{2p} \right\}^{1/p},
 \end{equation*}
 where the subscript $a$ represents algebraic extension. Two immediate special cases are 
 \begin{align}
 \Delta_{a}(X, p=1) & = \|V(\mathbf{Z})\|_{1} = 2H_{2}(X),\\
  \Delta_{a}(X, p=\infty) & =  \|V(\mathbf{Z})\|_{\infty} = \max_{1\leq k\leq K} (\pi_{k}-\pi^{2}_{k}).
 \end{align}
 
Equation 4 indicates that the Gini index (or R\'{e}nyi entropy) aligns with the $L_{1}$-norm of $V(\mathbf{Z})$. It is also noteworthy that $\Delta_{a}^{p}(X, p)$ is a special case of the more general framework $\Delta_{g}(X,~w,~l,~p)$ when $l = p$ and $w(\pi_{k}) = \pi_{k}^{p}$. Therefore, in Section 3, we will focus on exploring the validity of the broader concept $\Delta_{g}(X,~w,~l,~p)$.
 
 \section{Validity and counterintuitive results}
 To assess the validity of distance variance and related measures for categorical variables, we suggest a set of axioms for spread measures based on majorization and Schur-concavity \cite{MO1979}. We begin with the notations. Let $\mathcal{S}^{d} = \{(x_{1},~...,~x_{d}),~ \min_{1\leq i\leq d} x_{i}\geq 0,~\sum_{i=1}^{d}x_{i} = 1\}$ be the simplex of dimension $d$. Majorization is a concept used to describe how spread out a discrete probability distribution is. For $\pi,~\pi'\in S^{K}$, $\pi$ is said to be majorized by $\pi'$, denoted by $\pi\prec\pi'$, if the following condition is satisfied
\begin{equation*}
\sum_{k=i}^{K} \pi_{(k)} \leq \sum_{k=i}^{K} \pi'_{(k)}, ~ i = 2, ..., K, 
\end{equation*}
where $\pi_{(1)} \leq \pi_{(2)} \leq ... \leq \pi_{(K)}$ and $\pi'_{(1)} \leq \pi'_{(2)} \leq ... \leq \pi'_{(K)}$ represent the components of $\pi$ and $\pi'$ sorted in ascending order. We say $\pi$ is strictly majorized by $\pi'$, denoted by $\pi\prec_{s}\pi'$, if there exists at least one index $i$ such that $\sum_{k=i}^{K} \pi_{(k)} < \sum_{k=i}^{K} \pi'_{(k)}$.

This definition extends to vectors of different lengths by appending zeros to the shorter vector. An immediate observation is 
\begin{equation*}
(\frac{1}{K},~...,~\frac{1}{K})\prec (\frac{1}{K-1},~...,~\frac{1}{K-1},~0)\prec ... (\frac{1}{2},~\frac{1}{2},~0,~...,~0)\prec (1,~0,~...,~0).
\end{equation*}

To illustrate majorization further, consider two examples:
\begin{Exa}
$\left( 1/3,~ 1/3,~ 1/3 \right) \prec \left( 1/2,~ 1/2,~ 1/4 \right)$ because $1/3 < 1/2$ and $1/3 + 1/3 < 1/2 + 1/4$. The second distribution has higher probability mass concentrated in two categories, indicating less spread.
\end{Exa}
\begin{Exa}
$\left( 1/6,~1/6,~2/6,~2/6 \right) \prec \left( 1/10,~ 2/10,~ 3/10,~4/10 \right)$ because $2/6< 4/10$, $2/6 + 2/6 < 3/10 + 4/10$, and $1/6 + 2/6 + 2/6 < 2/10 + 3/10 + 4/10$. Again, the second distribution has a more concentrated probability mass, demonstrating less spread.
\end{Exa}

We propose a set of axioms for spread measures, denoted by $\Delta(\pi)$ or $\Delta(X)$ for a categorical variable $X$ with probability distribution $\pi$. These axioms aim to capture the concept of how spread out a distribution is across its categories.
\begin{itemize}
\item[(A1)] $\Delta(\pi)\geq 0$, where the equality holds if and only if there exists a single category $k$ such that $\pi_{k} = 1$. 
\item[(A2)] $\Delta(\pi)$ is continuously differentiable. 
\item[(A3)] If $\pi\prec_{s}\pi'$, $\Delta(\pi) > \Delta(\pi')$. 
\end{itemize}

\begin{Rem}
In the terminology of Marshall \& Olkin (1979), a functional satisfying A3 is said to be strictly Schur-concave (strictly increases as $\pi$ becomes more uniform). For continuously differentiable functions (as assumed in A2), strict Schur-covavity can be verified using partial derivatives (Marshall \& Olkin, 1979, Theorem A.4.b, page 85). To be specific, let $\Delta_{(k)} = \partial \Delta/ \partial \pi_{k}$ and $\Delta_{(k,k')} = \partial^{2} \Delta/ \partial \pi_{k}\partial \pi_{k'}$ be the first- and second-order partial derivatives of $\Delta$, necessary and sufficient conditions for $\Delta$ to be strictly Schur-concave are (1) $\Delta$ is twice differentiable and symmetric (2) $\Delta_{(k)}(\pi) > \Delta_{(k')}(\pi)$ for $\pi_{k} < \pi_{k'}$ (3) $\Delta_{(k)}(\pi) = \Delta_{(k')}(\pi)$ implies $\Delta_{(k,k)}(\pi) + \Delta_{(k',k')}(\pi) - \Delta_{(k,k')}(\pi) - \Delta_{(k',k)}(\pi) <0$.
\end{Rem}

\begin{Rem}
None of the proposed axioms necessarily imply the others. For example, A3 (Schur-concavity) in general does not guarantee A2 (continuous differentiability).
\end{Rem}

\begin{Rem}
Some existing spread measures, like D-dispersion (Gilula \& Haberman, 1995, Definition 1), incorporate symmetry as an axiom. In our axioms, A3 inherently implies symmetry. Let $\pi_{perm}$ be a permutation of $\pi$. As $\Delta$ is Schur-concave and $\pi\prec\pi_{perm}$, we have $\Delta(\pi) \geq \Delta(\pi_{perm})$. On the other hand, as $\pi_{perm}\prec\pi$, we have $\Delta(\pi_{perm}) \geq \Delta(\pi)$, therefore $\Delta(\pi_{perm}) = \Delta(\pi)$.
\end{Rem}

\begin{Rem}
Our axioms differ from the existing ones. For instance, A3 implies Shannon's second axiom regarding monotonicity (to be specific, $\Delta(1/K,~...,~1/K)$ is a monotonically increasing function of the dimension $K$). However, a measure satisfying A1-A3 might not satisfy Shannon's third axiom regarding additivity
\begin{equation*}
 \Delta(\pi_{1},~...,~\pi_{k-1},~t\pi_{k},~(1-t)\pi_{k},~\pi_{k+1},~...,~\pi_{K}) = \Delta(\pi) + \pi_{k} \Delta(t,~1-t),
\end{equation*}
for any $\pi_{k}\in[0,~1]$, any $t\in[0,~1]$ and any positive integer $K$. In his seminal paper, Shannon elegantly showed that the entropy function $\Delta(\pi) = -\sum_{k=1}^{K}\pi_{k}\log \pi_{k}$ is the only function that fulfills all his axioms. Additionally, A2 is a stronger condition compared to the continuity required by D-dispersion. However, continuous differentiability is necessary for verifying strict Schur-concavity. Finally, a D-dispersion that is also strictly concave satisfies strict Schur-concavity.
\end{Rem}

We now examine the validity of some well-known measures. First, it is straightforward to verify that Shannon entropy, extropy and Tsallis entropy $(m>1)$ all satisfy the three axioms. Taking Tsallis entropy as an example, A1 is satisfied as $\sum_{k=1}^{K} \pi^{m}_{k} \leq \sum_{k=1}^{K} \pi_{k}$ (equality holds only when $K=1$). Additionally, $H_{m}(\pi)$ is symmetric and twice differentiable with partial derivative $H_{m(k)}(\pi) = -m\pi_{k}^{m-1}/(m-1)$, therefore $H_{m(k)}(\pi)>H_{m(k')}(\pi)$ for $\pi_{k} < \pi_{k'}$. Finally, the second-order partial derivatives are $H_{m(k,~k)}(\pi) = -m\pi_{k}^{m-2}$ and $H_{m(k,~k')}(\pi) = 0$, therefore $H_{m(k,k)}(\pi) + H_{m(k',k')}(\pi) - H_{m(k,k')}(\pi) - H_{m(k',k)}(\pi) <0$. As a special case of Tsallis entropy when $m=2$, Gini index also qualifies as a valid measure.

For $K=2$, distance variance can be expressed as $2\sqrt{2}\pi(1-\pi)$ (same as variance up to a constant), which satisfies the three axioms. However, for $K\geq 3$, distance variance does not satisfy A3, suggesting that it is not a valid measure of spread for general categorical variables. Figures 2 and 3 illustrate some counterintuitive results. Figure 2 demonstrates that for equally probable distributions, $\pi_{k} = 1/K$, distance variance peaks at $K=2$ and progressively decreases to 0 with increasing $K$. Figure 3 presents another example where a distribution with equal probabilities (case 2) exhibits lower variability according to distance variance compared to a distribution with unequal probabilities (case 1). 

\begin{figure}[!htbp]
\begin{center}
\includegraphics[scale=0.58]{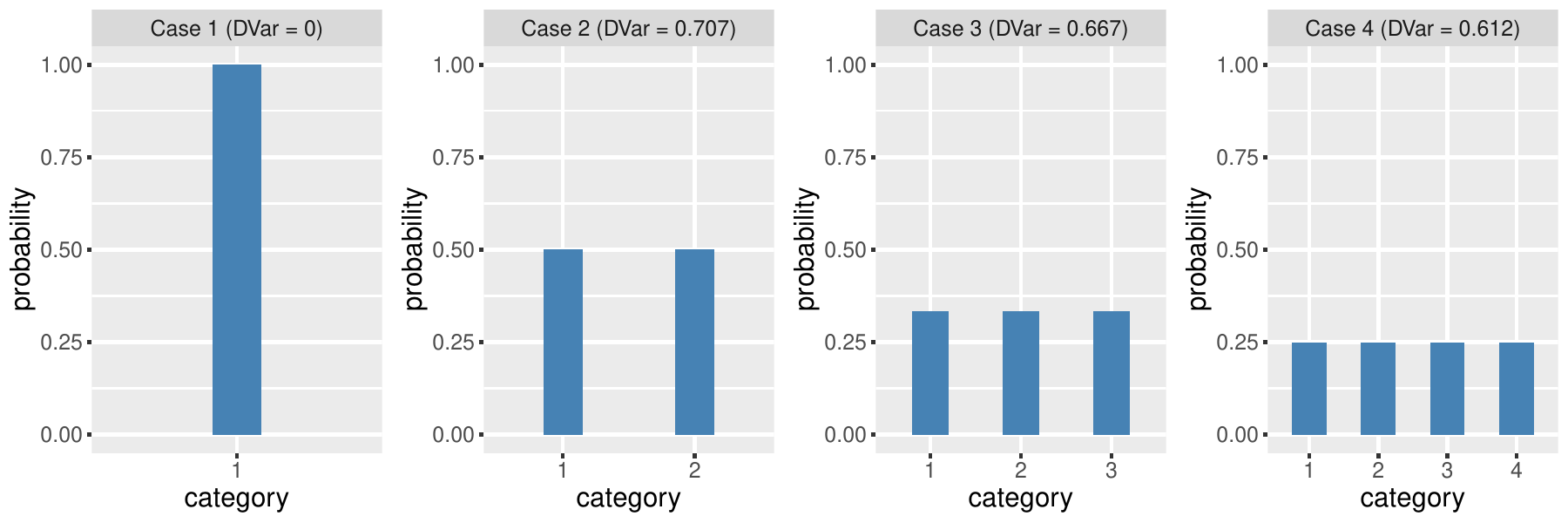}
\end{center}
\caption{Distance variance of equal-probability distributions with $K=1,~,2~,3,~4$.}
\end{figure}

\begin{figure}[!htbp]
\begin{center}
\includegraphics[scale=0.5]{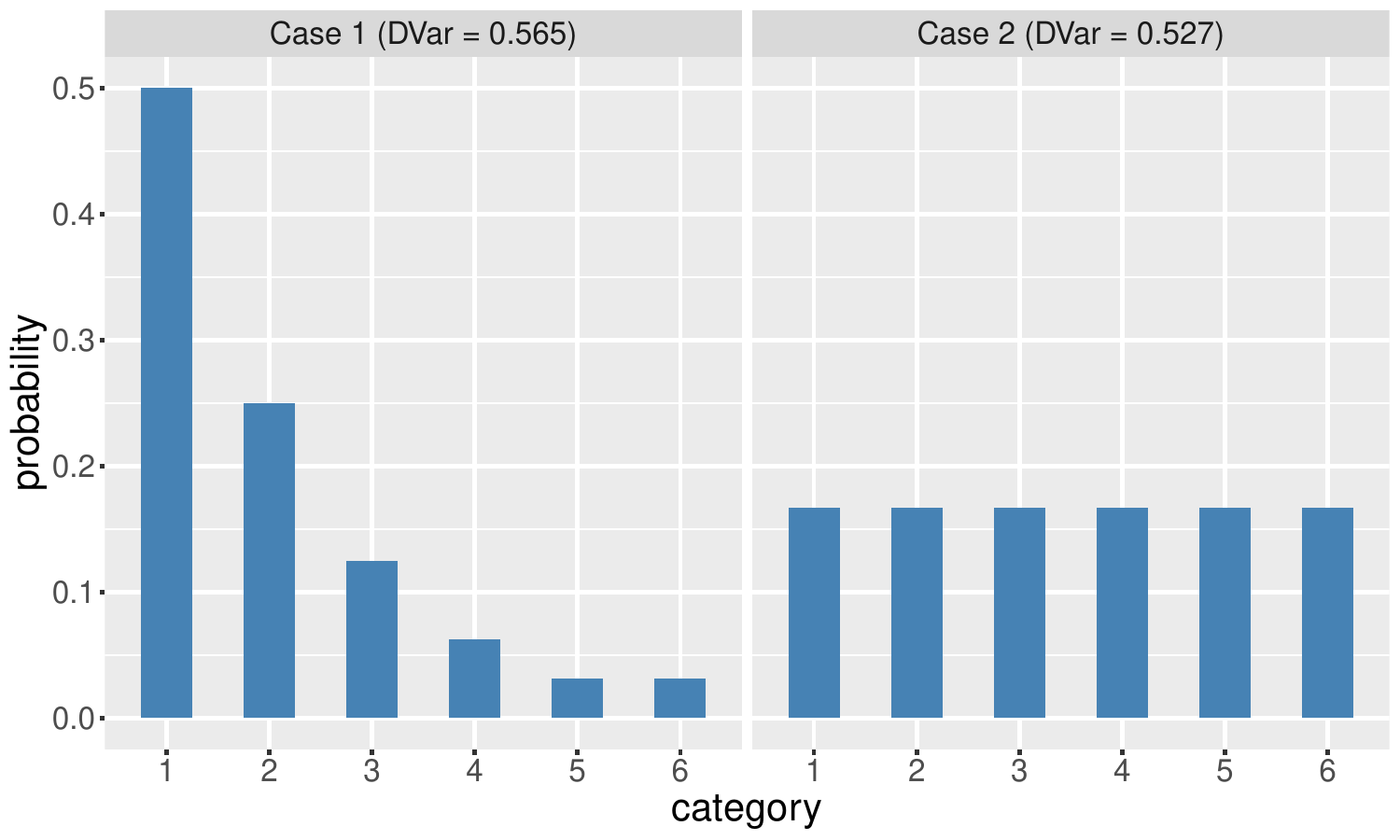}
\end{center}
\caption{Distance variance of two distributions ($K=6$): Case 1 - $(1/2,~1/4,~1/8,~1/16,~1/32,~1/32)$; Case 2 - equal probabilities.}
\end{figure}

Within the extended class $\Delta_{g}(X,~w,~l,~p)$, we find that when $p=1$ (average $L_1$ distance), it is easier to construct new spread measures that satisfy the three axioms. For $p\geq 2$, constructing valid measures becomes mathematically more challenging. For $p=1$, the following new metrics satisfy all three axioms. These metrics have simple expression and it is likely that many more such metrics exist
\begin{itemize}
\item $\sum_{k=1}^{K}\exp{(\pi_{k})}(1-\pi_{k})$, i.e., $l=1$, $w(\pi_{k}) = \exp{(\pi_{k})}$.
\item $\sum_{k=1}^{K}\sin(\pi_{k})(1-\pi_{k})$, i.e., $l = 1$, $w(\pi_{k}) = \sin(\pi_{k})$. 
\end{itemize}

\section{Discussion and conclusions}
Edelmann et al. (2020) established distance variance, a quantity derived from distance correlation analysis, as a valuable spread measure for continuous and binary data. This work explores its potential as a universal metric by investigating its applicability to nominal variables with more than two categories. Our findings reveal that, unfortunately, distance variance in its current form is not a valid measure of spread for general categorical variables. Misusing it can lead to counterintuitive results. Therefore, when applied to categorical data, distance variance should not be interpreted as spread, but rather as an average weighted distance to the geometric center, or the $L_2$-norm (Frobenius norm) of the variance-covariance matrix based on the one-hot encoded representation of the categorical variable.

This work focuses on distance variance based on binary distances. A natural question arises: could alternative, non-binary distance metrics based on category probabilities make distance variance a valid spread measure? van de Velden et al. (2023) summarized several such metrics for categorical variables that could be explored within the framework of distance variance. For instance, Lin (1998) proposed a metric that gives more weight to matches on frequent values and lower weight to mismatches on infrequent values:
\begin{equation*}
d(k,~k') = \frac{\log(\pi_{k}) + \log(\pi_{k'}) - 2\log(\pi_{k} + \pi_{k'})}{2\log(\pi_{k} + \pi_{k'})}.
\end{equation*}
Similarly, Boriah et al. (2008) introduced the inverse occurrence frequency (IOF) distance, which increases with the co-occurrence frequency of categories, akin to the inverse document frequency concept in information retrieval. Replacing binary distances with such frequency-dependent metrics might lead to a new form of distance variance that satisfies our proposed axioms for spread. This remains an open question for future investigation.

\section*{Acknowledgement}
\noindent
The work was supported by an NSF DBI Biology Integration Institute (BII) grant (award no. 2119968; PI-Alrubaye).

 \section*{Competing Interests}
\noindent
The author has declared that no competing interests exist.

\section*{Appendix}
\subsection*{Derivation of $\mathcal{V}(X)$ for arbitrary distances}
Define $d(x_{1},~x_{2}) = c_{1}$ if $x_{1} = x_{2}$ and $d(x_{1},~x_{2}) = c_{2}$ if $x_{1} \neq x_{2}$, we have
\begin{align*}
E\left[d(X_{1},~X_{2})\right] & = c_{1}P(X_{1} = X_{2}) + c_{2}P(X_{1} \neq X_{2}) \\
& = c_{1}\sum_{k=1}^{K}\pi^{2}_{k} + c_{2}\left( 1- \sum_{k=1}^{K}\pi^{2}_{k} \right), \\
E\left[d^{2}(X_{1},~X_{2})\right] & = c^{2}_{1}P(X_{1} = X_{2}) + c^{2}_{2}P(X_{1} \neq X_{2}) \\
& = c^{2}_{1}\sum_{k=1}^{K}\pi^{2}_{k} + c^{2}_{2}\left( 1- \sum_{k=1}^{K}\pi^{2}_{k} \right),
\end{align*}
and
\begin{align*}
E\left[d(X_{1},~X_{2})d(X_{1},~X_{3})\right] & = c^{2}_{1}P(X_{1} = X_{2} = X_{3}) + c^{2}_{2}P(X_{1}\neq X_{2},~X_{1}\neq X_{3}) + 2c_{1}c_{2}P(X_{1} = X_{2},~X_{1}\neq X_{3})  \\
& = c^{2}_{1}\sum_{k=1}^{K}\pi^{3}_{k} + c^{2}_{2}\sum_{k=1}^{K}\pi_{k}(1-\pi_{k})^{2} + c_{1}c_{2}\left[1-\sum_{k=1}^{K}\pi^{3}_{k}-\sum_{k=1}^{K}\pi_{k}(1-\pi_{k})^{2}\right].
\end{align*}
Summarizing the results above, the distance variance of $X$ can be expressed as 
\begin{equation*}
\mathcal{V}(X,~c_{1},~c_{2}) = |c_{1}-c_{2}| \left[ \left( \sum_{k=1}^{K}\pi_{k}^2 \right)^2 + \sum_{k=1}^{K}\pi_{k}^2 - 2\sum_{k=1}^{K}\pi_{k}^3 \right]^{\frac{1}{2}}.
\end{equation*}

\end{document}